\documentclass[singlespacing]{elsart}

\usepackage{graphicx}

\usepackage{amssymb}
\journal{Computational and Theoretical Chemistry}
\begin{document}

\begin{frontmatter}

\title{Exactly solvable double-well potential in Schr\"odinger equation for inversion mode of phosphine molecule}

\author{ A.E. Sitnitsky},
\ead{sitnitsky@kibb.knc.ru}

\address{Kazan Institute of Biochemistry and Biophysics, FRC Kazan Scientific Center of RAS, P.O.B. 30,
420111, Russian Federation. e-mail: sitnitsky@kibb.knc.ru}

\begin{abstract}
The reduced mass of the effective quantum particle for the inversion mode $\nu_2$ of phosphine molecule ${\rm{PH_3}}$ is known to be a position dependent one. In the present article the inversion spectrum of ${\rm{PH_3}}$ is considered with the help of the Schr\"odinger equation (SE) with position dependent mass and corresponding modified double-well potential. The SE is shown to be exactly solvable. The results are used for the analysis of the pertinent experimental data available in literature. We are based on the reliable value $\nu_2=E_2-E_0=992.1\ {\rm cm^{-1}}$ ($2\nu_2=E_4-E_0=1972.5\ {\rm cm^{-1}}$; $3\nu_2=E_6-E_0=2940.8\ {\rm cm^{-1}}$; $4\nu_2=E_8-E_0=3895.9\ {\rm cm^{-1}}$) obtained by \v Spirko et al. Also we use the value for the barrier height $E_b=12300 \ {\rm cm^{-1}}$ that seems to be commonly accepted at present and the hypothetical value for the energy splitting of the 11-th doublet of the $10\nu_2$ band $s_{10}=E_{21}-E_{20}\approx 7.2\ {\rm cm^{-1}}$ suggested in the literature. Definite predictions are derived for the energy splitting of the 4-th doublet of the $3\nu_2$ band $s_3=E_{7}-E_{6}$ that is a test one for the observation in the experiment of Okuda et al. SE with position dependent mass provides self-consistently the required values of $\{\nu_2; E_b;s_{10}\}$ yielding $s_3= 6.21\cdot 10^{-12}\ {\rm cm^{-1}}$.
\end{abstract}

\end{frontmatter}

\section{Introduction}
Phosphine molecule ${\rm{PH_3}}$ (see \cite{Spi78}, \cite{Spi86}, \cite{Sch92}, \cite{Sou15}, \cite{Zu13}, \cite{Sou13}, \cite{Dev14}, \cite{Sou16}, \cite{Sou14}, \cite{Sme97}, \cite{Nik09}, \cite{Nik14}, \cite{Cre98}, \cite{Yur06}, \cite{Mul13}, \cite{Oku18} and refs. therein) is an interesting counterpart of the famous (for its distinguished role in the invention of the first maser) ammonia one ${\rm{NH_3}}$ \cite{Tow55}, \cite{Laa09}, \cite{Rus97}, \cite{Cro05}. The steady interest to ${\rm{PH_3}}$ is stimulated by its importance for astrophysics (spectroscopic studies of giant-planets and cool stars) \cite{Sou15}. It has been increased by recent finding of this molecule in Venus atmosphere \cite{Gre20} with speculations about its anaerobic microbial life origin. For the ammonia molecule a rather simple energy spectrum of the inversion mode is observed with a pair of doublets below the barrier top and a pronounced ground state splitting $E_1-E_0=0.79\ {\rm cm^{-1}}$ \cite{Laa09}. The change of nitrogen ${\rm{N}}$ by phosphorus ${\rm{P}}$ drastically alters the energy levels structure and makes it to be qualitatively different. There is an unknown number of doublets below the barrier top with still experimentally indiscernible ground state splitting \cite{Sou15}, \cite{Sou16}. Along with the problems for experimental studies of the symmetric bending mode $\nu_2$ (umbrella mode or inversion mode) in ${\rm{PH_3}}$ its description is a challenge to theory. Up to now the theoretically predicted inversion splitting for various doublets $n\nu_2$ is not confirmed by the experiment (e.g., that for $3\nu_2$ band $s_3=E_{7}-E_{6}\approx 1\cdot 10^{-5}\ {\rm cm^{-1}}$ predicted in \cite{Sou15}, \cite{Sou16} is not observed at the indicated frequency \cite{Oku18}). Also there are no theoretical approaches to the calculation of the ground state splitting for ${\rm{PH_3}}$ (only the estimates like $\leq 10^{-10}\ {\rm cm^{-1}}$ are available \cite{Sou16}).

In our opinion there is an urgent need in a suitable double-well potential (DWP) for this mode of ${\rm{PH_3}}$. A crucial requirement is that the corresponding Schr\"odinger equation (SE) can be satisfactorily solved. This solution should be convenient to enable one easy fitting the energy levels and doublets splitting. DWPs for which SE can be solved only numerically or by approximate methods like, e.g., quasi-classical (WKB) procedure do not satisfy this requirement. Besides WKB fails for energy levels near the barrier top \cite{Lan74} that is well corroborated by cases where an exact analytic solution is available \cite{Sit18}. Thus the matter of the convenience in usage and the ability to provide easy scanning the parameter space of the model under consideration becomes of utmost importance. In this regard exact analytic solutions in quantum mechanics if they are feasible always take a superior position. One can conclude that SE with the desired DWP for the inversion mode of ${\rm{PH_3}}$ should be amenable to exact analytic solution via well studied special functions. Moreover for the solution to be of real practical use it is necessary that these functions are implemented in commonly accessible mathematical software packages like, e.g., {\sl {Mathematica}} or {\sl {Maple}}. During last years a number of DWPs for SE with constant mass suitable for problems of chemistry (infinite at the boundaries of the interval for the spatial variable) was suggested. For them analytic solutions were obtained via the confluent Heun's function (implemented in {\sl {Maple}}) \cite{Sit17}, \cite{Sit171}, \cite{Don18}, \cite{Don181}, \cite{Don182}, \cite{Don183}, \cite{Don19}, \cite{Don191}, \cite{Don192}, \cite{Don20}  and the spheroidal function (implemented in {\sl {Mathematica}}) \cite{Sit171}, \cite{Sit18}. Unfortunately they are not suitable for the problem of phosphine because in this case we deal with a position dependent mass \cite{Aqu98}, \cite{For12} (see below).

It is commonly accepted that the inversion mode of a pyramidal molecule like ${\rm{NH_3}}$, ${\rm{PH_3}}$, etc. can be conceived as a movement of a quantum particle. This particle is effective in the sense that its reduced mass (specifying further ${\rm{P}}$ atom) does not coincide either with that of ${\rm{P}}$ or ${\rm{H}}$ atoms. The pertinent degree of freedom for the effective particle corresponds to the motion of the $P$ atom relative to the $H_3$ symmetrical triangle. The particle moves under the influence of DWP along a spatial coordinate. DWP results from the Coulomb repulsion between the phosphorus nucleus and the three protons. As a matter of fact three hydrogen atoms simultaneously tunnel from one side of the phosphorus to the other though the ${\rm{P}}$ atom also slightly adjusts its position to provide that the center of mass remains constant \cite{Cro05}. However even in the early articles it was anticipated that the picture with a constant mass of the effective particle for the inversion mode of pyramidal molecules may be oversimplified (see \cite{Tow55} and refs. therein). Thus a concept of the position dependent mass for the effective particle was introduced. It was considered self-consistently for the first time by the authors of \cite{Aqu98}. However for ${\rm{NH_3}}$ treated in \cite{Aqu98} this fact is of no crucial importance and the subject did not receive due attention. Besides it should be stressed that the whole analysis in \cite{Aqu98} is carried out within the framework of classical mechanics that is certainly inadequate for the quantum tunneling process. For this reason we consider such analysis merely as suggestive arguments. Instead of it we introduce a purely empirical position dependent mass for the purpose of the phenomenological description of available literature data on $\nu_2$. We do not try to provide any stringent justification of our position dependent mass function. We suggest it as a phenomenological relationship and the aim of the present article is to show that it yields the required description of the inversion energy spectrum of ${\rm{PH_3}}$. However in Appendix we briefly outline how our position dependent mass function can be obtained within the formalism of \cite{Aqu98}.

The direct problem of constructing a modified potential from a given position dependent mass is well studied (see, e.g., \cite{Don07ba}, \cite{Don04aa}, \cite{Don07aa}, \cite{Akt08}, \cite{Don16aa}, \cite{Pe17} and refs. therein). The crucial fact is that a suitable transformation of SE with position dependent mass to that with constant mass does not alter the spectrum of the energy levels. In the present article we apply this technique to reduce SE with the appropriate (for phosphine) position dependent mass to that with the constant mass and the trigonometric DWP (TDWP). SE with the constant mass and TDWP fails to describe available data on the inversion of ${\rm{PH_3}}$ (see Sec.4). Thus we have to resort the concept of the position dependent mass. In fact we solve the inverse problem. We know how we want to modify TDWP in order the modified DWP describes the data and we seek a suitable position dependent mass function that provides such transformation. SE with TDWP is exactly solvable \cite{Sit18} and the aim of the present article is to show that it can be used (after the above mentioned transformation) for the calculation of the energy levels structure for the inversion mode of ${\rm{PH_3}}$. The merit of TDWP is in its convenience because SE with it is solved via the spheroidal function (SF) \cite{Kom76} implemented in {\sl {Mathematica}} along with its spectrum of eigenvalues. As a result the calculation of the energy levels is a very simple procedure. This circumstance makes TDWP to be an extremely convenient and efficient tool for usage. Earlier TDWP was applied to an asymmetric hydrogen bond in ${\rm{KHCO_3}}$ \cite{Sit17}, inversion of an ammonia molecule ${\rm{NH_3}}$ \cite{Sit171}, \cite{Cai20}, ring-puckering vibration in 1,3-dioxole and  2,3-dihydrofuran \cite{Sit18}, the calculations of the polarizability of hydrogen bonds in chromous acid (CrOOH) and potassium dihydrogen phosphate (${\rm{KH_2PO_4}}$) \cite{Sit19} and the calculation of IR absorption intensities for a hydrogen bond in the Zundel ion ${\rm{H_5O_2^{+}}}$ (oxonium hydrate) \cite{Sit20}. In the present article we construct with the help of TDWP a modified phenomenological DWP for the exactly solvable SE with the position-dependent mass that yields the description of the energy levels structure for the inversion mode of ${\rm{PH_3}}$.

The parameters of phenomenological DWPs are usually determined determined either by comparing with the results of full-fledged quantum chemical {\it ab initio} calculations of the potential energy surface (PES) or from IR spectroscopy data. To make further use of the results of quantum chemical calculations one inevitably has to approximate one-dimensional cross-sections of PES by some DWPs that are usually high-degree polynomials (see, e.g., \cite{Rus97}). SE for such DWPs can be treated only numerically. As mentioned above in this regard those rare cases of DWPs are of particular interest for which exact analytic solutions via well studied special functions are feasible. Our TDWP is one of them. It is shown to model PES for hydrogen bonds \cite{Sit20} but in the present article we do not directly follow along this line in dealing with ${\rm{PH_3}}$. We make use of the combined information available in literature (the barrier height from PES \cite{Sch92} along with data from the theoretical models supported by experimental results \cite{Spi78}, \cite{Spi86}). Our approach makes the calculation of the energy levels structure to be a routine (at a click) procedure. As a result it enables one prompt scanning the parameter space of the model. Also it provides convenient calculation of various values such as dipole matrix elements between the wave functions of different energy levels and absorption intensities \cite{Sit20}. We show that our phenomenological DWP yields good description of available data for ${\rm{PH_3}}$. However such data are rather scarce and uncertain at present. Thus our approach should be conceived as an additional efficient tool for quick fitting and interpreting the newly appearing information in the toolkit of practitioners dealing with IR spectroscopy of ${\rm{PH_3}}$.

The structure of the article is as follows. In Sec.2 the solution for SE with the position dependent mass is constructed. In Sec.3 the position dependent mass pertinent for the problem of phosphine and the resultant effective potential are considered. In Sec.4 the results are discussed and the conclusions are summarized. In Appendix some technical information is presented.

\section{Schr\"odinger equation with position dependent mass}
We write the position dependent mass in the form
\begin{equation}
\label{eq1} M(X)=M_0f(X);\ \ \ \ \ \ \ M_0=const\ \frac{3m_Hm_P}{3m_H+m_P}; \ \ \ \ \ \ \ -L \leq X \leq L
\end{equation}
where $const$ is a constant to be determined later, $f(X)$ is a function to be obtained from fitting the pertinent IR spectroscopy data for ${\rm{PH_3}}$ and
$L$ is a characteristic length to be defined later (see Appendix). The dimensional SE for a wave function $\Psi (X)$ is \cite{Don07ba}
\begin{equation}
\label{eq2} \left\{-\frac{\hbar^2}{2M_0}\frac{d }{dX}\frac{1}{f\left(X\right)}\frac{d }{dX}+\Omega\left(X\right)-E \right\}\Psi (X)=0
\end{equation}
where $\Omega\left(X\right)$ is a DWP to be defined below (see Appendix). We introduce the dimensionless values for the distance $x$, the corresponding modified
 DWP $W(x)$ and the dimensionless energy $\epsilon$
\begin{equation}
\label{eq3} x=\frac{X}{L};\ \ \ \ \ \ \ \ \ \ W(x)=\frac{2M_0L^2}{\hbar^2}\Omega\left(X\right);\ \ \ \ \ \ \ \ \ \ \epsilon=\frac{2M_0L^2E}{\hbar^2}
\end{equation}
Dimensionless SE takes the form
\begin{equation}
\label{eq4} \left\{-\frac{d }{dx}\frac{1}{f(x)}\frac{d }{dx}+W\left(x\right)-\epsilon\right\}\Psi (x)=0
\end{equation}
For the given position dependent mass function in the dimensionless designations $f(x)$ we follow the well known procedure (see, e.g., \cite{Don07ba}, \cite{Don04aa}, \cite{Don07aa}, \cite{Akt08}, \cite{Don16aa}, \cite{Pe17} and refs. therein). We introduce a new variable $y(x)$ defined by the requirement
\begin{equation}
\label{eq5} y(x)=\int^{x}_0 dz\ \sqrt{f(z)}\ \ \ \ \ \ \ \ \ \ \ \ \ \ \ \ \ \ \ \ \ \ \ \ \ \ \ -\pi/2\leq y \leq \pi/2
\end{equation}
so that
\begin{equation}
\label{eq6} y'_x(x)=f^{1/2}(x)\ \ \ \ \ \ \ \ \ \ \ \ \ \ \ \ \ \ \ \ \ \ \ \ \ \ \ \ \ \ \ \ \ \ x'_y(x)=f^{-1/2}(x)
\end{equation}
Thus we have for the function $y(x)$ the requirement
\begin{equation}
\label{eq7} y\left(\pm 1\right)=\pm \pi/2
\end{equation}
We designate the function
\begin{equation}
\label{eq8} g(y)=-\frac{1}{2}\left(\frac{f'_x(x)}{f^{3/2}(x)}\right)(y)=\left(\frac{1}{\sqrt f}\right)'_x(y);\ \ \ \ \ \ \ \ \ g(x)=\left(\frac{1}{\sqrt f}\right)'_x
\end{equation}
SE (\ref{eq4}) takes the form
\begin{equation}
\label{eq9} \Psi''_{yy} (y)+g(y)\Psi'_{y} (y)+\left[\epsilon-W(y)\right]\Psi (y)=0
\end{equation}
We introduce a new function
\begin{equation}
\label{eq10} \varphi(y)=\Psi(y)\exp\left(\frac{1}{2}\int^{y}dz\ g(z)\right)
\end{equation}
The transformed SE for $\varphi(y)$ is
\begin{equation}
\label{eq11} \varphi''_{yy} (y)+\left[\epsilon-W(y)-\frac{1}{4}g^2(y)-\frac{1}{2}g'_y(y)\right]\varphi (y)=0
\end{equation}
This is an ordinary SE for a quantum particle with position-independent mass $M_0$ in an effective potential $W(y)+\frac{1}{4}g^2+\frac{1}{2}g'_y(y)$. We require this potential to be TDWP $U(y)$ studied in \cite{Sit18}
\begin{equation}
\label{eq12} U(y)=\frac{B\left[\cos\left(\Delta/2\right)\right]^4}{\left(1-\left[\cos\left(\Delta/2\right)\right]^2\right)^2}\ \tan^2 y-\frac{B}{\left(1-\left[\cos\left(\Delta/2\right)\right]^2\right)^2}\sin^2 y
\end{equation}
Here $-\pi/2\leq y \leq \pi/2$ and we denote the barrier height of TDWP $B=-U\left(y_{min}\right)$ and its barrier width $\Delta=y_{min}^{(1)}-y_{min}^{(2)}$. The example of TDWP is presented in Fig.1 as the dashed line.

Further it is convenient to denote
\begin{equation}
\label{eq13}\frac{B\left[\cos\left(\Delta/2\right)\right]^4}{\left(1-\left[\cos\left(\Delta/2\right)\right]^2\right)^2}=m^2-\frac{1}{4}\ \ \ \ \ \ \ \ \ \ \ \ \ \ \ \ \frac{B}{\left(1-\left[\cos\left(\Delta/2\right)\right]^2\right)^2}=p^2
\end{equation}
Here $p$ is a real number and we further choose $B$ and $\Delta$ so that $m$ is an integer number. Inversely we obtain
\begin{equation}
\label{eq14}
\Delta(m,p)=2\arccos\left(\frac{m^2-\frac{1}{4}}{p^2}\right)^{1/4}
\end{equation}
\begin{equation}
\label{eq15}
B(m,p)=\left(\sqrt {m^2-\frac{1}{4}}-p\right)^2
\end{equation}
SE (\ref{eq11}) with the potential (\ref{eq12}) takes the form
\begin{equation}
\label{eq16}\varphi''_{yy} (y)+\left[\epsilon-\left(m^2-\frac{1}{4}\right)\tan^2 y +p^2 \sin^2 y\right]\varphi(y)=0
\end{equation}
The solution of (\ref{eq16}) is \cite{Sit18}
\begin{equation}
\label{eq17} \varphi_q (y)=\cos^{1/2} y\ \bar S_{m(q+m)}\left(p;\sin y\right)
\end{equation}
Here $q=0,1,2,...$ and $\bar S_{m(q+m)}\left(p;s\right)$ in the designations of \cite{Kom76} is the normalized angular prolate SF. This function is implemented in {\sl {Mathematica}} as $\rm{SpheroidalPS}[(q+m),m,ip,s]$. It should be stressed that the latter function is not normalized. The energy levels are determined by the relationship
\begin{equation}
\label{eq18}
\epsilon_q=\lambda_{m(q+m)}\left(p\right)+\frac{1}{2}-m^2-p^2
\end{equation}
Here $\lambda_{m(q+m)}\left(p\right)$ in the designations of \cite{Kom76} is the spectrum of eigenvalues for $\bar S_{m(q+m)}\left(p;s\right)$. It is implemented in {\sl {Mathematica}} as\\ $\lambda_{m(q+m)}\left(p\right)\equiv \rm{SpheroidalEigenvalue}[(q+m),m,ip]$.

Thus our modified DWP is defined as
\begin{equation}
\label{eq19} W(y)=\left(m^2-\frac{1}{4}\right)\tan^2 y-p^2 \sin^2 y-\frac{1}{4}g^2(y)-\frac{1}{2}g'_y(y)
\end{equation}
For $g'_y$ we have
\begin{equation}
\label{eq20} g'_y=g'_x x'_y=g'_x \frac{1}{\sqrt f}
\end{equation}
We write the required DWP in a convenient form for further analysis
\begin{equation}
\label{eq21} W(x)=\left(m^2-\frac{1}{4}\right)\tan^2 y(x)-p^2 \sin^2 y(x)+\frac{g^2(x)}{4}-\left[\frac{g^2(x)}{2}+\frac{g'_x(x)}{2\sqrt {f(x)}}\right]
\end{equation}

\section{Position dependent mass and effective potential}
Our purpose that will be discussed in the next Sec. is to increase the barrier height (the central part of TDWP) without touching upon the energy levels structure and considerable increase of the wings (noncentral parts) of the DWP. For this reason we require
\begin{equation}
\label{eq22} -\left[\frac{g^2(x)}{2}+\frac{g'_x(x)}{2\sqrt {f(x)}}\right]=\omega(x)
\end{equation}
where we choose $\omega(x)$ to be a rapidly decreasing function from the center
\begin{equation}
\label{eq23} \omega(x)=\frac{a}{1+b x^2}
\end{equation}
Taking into account (\ref{eq8}) it can be written as an ordinary differential equation for the function $f(x)$
\begin{equation}
\label{eq24} f''_{xx}-\frac{2}{f}\left(f'_x\right)^2-4\omega(x)f^2=0
\end{equation}
that is of the type N6.52 from Ch.VI of \cite{Kam71}. Its solution for (\ref{eq23}) with the appropriate choice of the constants of integration is
\begin{equation}
\label{eq25} f(x)=\frac{\sqrt b}{4a}\left[\left(\arctan \sqrt b +\frac{\ln \left(1+b\right)}{2\sqrt b}\right)-\left(x \arctan \sqrt b x+\frac{\ln \left(1+b x^2\right)}{2\sqrt b}\right)\right]^{-1}
\end{equation}
The variable $y(x)$ satisfying (\ref{eq7}) is
\[
y(x)=C\ sgn\ x\int^{\mid x \mid}_0 dz\ \times
\]
\begin{equation}
\label{eq26} \left[\sqrt{\left(\arctan \sqrt b +\frac{\ln \left(1+b\right)}{2\sqrt b}\right)-\left(z \arctan \sqrt b z+\frac{\ln \left(1+b z^2\right)}{2\sqrt b}\right)}\right]^{-1}
\end{equation}
where
\[
C=\frac{\pi}{2}\Biggl\{\int^{1}_0 dz\ \times
\]
\begin{equation}
\label{eq27} \left[\sqrt{\left[\arctan \sqrt b +\frac{\ln \left(1+b\right)}{2\sqrt b}\right]-\left[z \arctan \sqrt b z+\frac{\ln \left(1+b z^2\right)}{2\sqrt b}\right]}\right]^{-1}\Biggr\}^{-1}
\end{equation}
Thus we have all terms to substitute in (\ref{eq21}). The resulting potential $W(x)$ is
 \begin{equation}
\label{eq28} W(x)=\left(m^2-\frac{1}{4}\right)\tan^2 y(x)-p^2 \sin^2 y(x)+\frac{1}{4}\left[\left(\frac{1}{\sqrt f}\right)'_x\right]^2+\frac{a}{1+b x^2}
\end{equation}
Further substitution of (\ref{eq25}) and (\ref{eq26}) in (\ref{eq28}) leads to a very cumbersome expression and we do not write it down here explicitly to save room. However it is easily programmed and the result is depicted in Fig.1. The energy levels $\epsilon_q$ for the wave function $\Psi_q (x)$ as the solution of SE (\ref{eq4}) with this DWP are given by (\ref{eq18}).
\begin{figure}
\begin{center}
\includegraphics* [width=\textwidth] {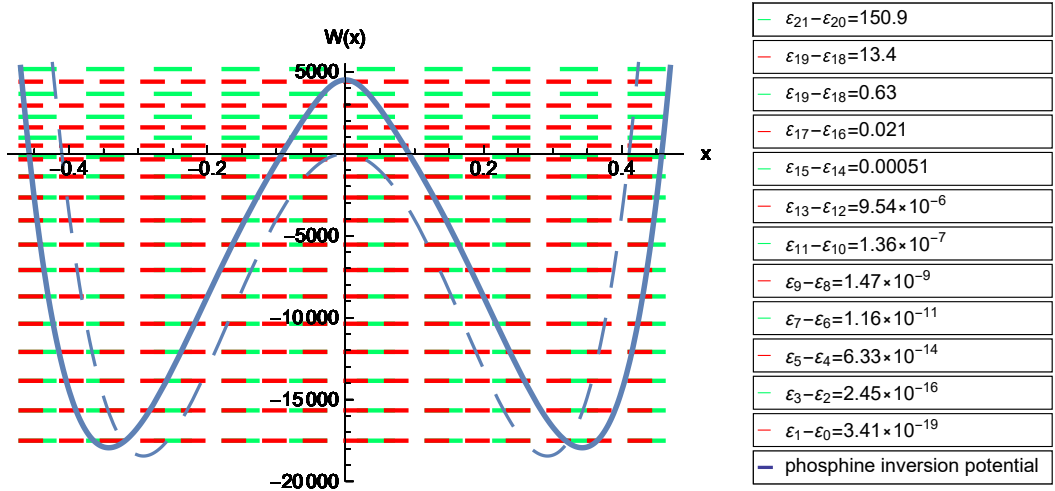}
\end{center}
\caption{The double-well potential (\ref{eq28}) at the values of the parameters $m=1500$, $p=1635.8222$, $a = 4500$ and $b = 150$. The parameters are chosen to describe the energy levels for the inversion mode $\nu_2$ of phosphine molecule. Experimental data are taken from \cite{Spi86} ($\nu_2=992.1\ {\rm cm^{-1}}$) along with literature data for the barrier height $E_b=12300 \ {\rm cm^{-1}}$ \cite{Sch92} and for the supposed splitting $E_{21}-E_{20}\approx 7.2\ {\rm cm^{-1}}$ for $10\nu_2$ \cite{Sou15}. The splitting of the ground state corresponds to $E_{1}-E_{0}=1.83\cdot 10^{-19}\ {\rm cm^{-1}}$ in dimensional units. The splitting for $3\nu_2$ corresponds to $s_3=E_{7}-E_{6}=6.21\cdot 10^{-12}\ {\rm cm^{-1}}$ (that was sought for in \cite{Oku18} as $E_{7}-E_{6}\approx 1\cdot 10^{-5}\ {\rm cm^{-1}}$ according to the prediction of \cite{Sou16}). The dashed line corresponds to the ordinary double-well potential (\ref{eq12}).} \label{Fig.1}
\end{figure}

\section{Results and discussion}
In \cite{Aqu98}, \cite{For12} the reduced mass of the effective quantum particle at the inversion of a pyramidal molecule (specifying here ${\rm{P}}$ atom although the authors of \cite{Aqu98}, \cite{For12} treat ${\rm{N}}$ one) is suggested to be a position dependent one and approximated as
\begin{equation}
\label{eq29} M(x)=\frac{3m_Hm_P}{3m_H+m_P}\left(1+\frac{3m_H+m_P}{m_P}\frac{x^2}{1-x^2}\right)
\end{equation}
The inversion coordinate in dimensionless form  $x=X/r_0$ corresponds to the current distance of the ${\rm{P}}$ atom left (or right) the ${\rm{H_3}}$ plane where $r_0$ is the dimensional distance from the ${\rm{P}}$ atom to one of the ${\rm{H}}$ atoms. As $X$ can not be greater than $r_0$ ($-r_0 \leq X \leq r_0 $)  then $ r_0$ is equivalent to our limit value $L$ in (\ref{eq1}) ($r_0 \equiv L$ in the designations of the present article). The distance $r$ from the ${\rm{P}}$ atom to one of the ${\rm{H}}$ atoms is assumed to remain constant during the inversion process $r=r_0$ and equals to its initial distance $r_0$. The formula (\ref{eq29}) is argued in \cite{Aqu98} to suit well for ${\rm{NH_3}}$ inversion energy levels structure but unfortunately it fails in our case of ${\rm{PH_3}}$. In contrast to our approach it yields undesirable considerable increase of the wings (noncentral parts) of the DWP. In the present article we release the above mentioned requirement that $r=r_0$ during the inversion. We use the phenomenological position dependent mass function $f(x)$ given by (\ref{eq25}). In Appendix we derive it within the framework of the formalism of \cite{Aqu98} under discarding the above requirement $r=r_0$ and a specific choice of $r(x)$. However we stress once more that such derivation with the help of the classical mechanics can not be considered as an adequate validation in our quantum tunneling problem. Instead of it we introduce (\ref{eq25}) as a purely phenomenological position dependent mass function suggested to describe pertinent experimental data for the inversion of ${\rm{PH_3}}$. Following \cite{Aqu98} we require that our $M_0$ in (\ref{eq1}) is defined by the requirement
\begin{equation}
\label{eq30} M(0)=M_0f(0)=\frac{3m_Hm_P}{3m_H+m_P}; \ \ \ \ \ \ \ \ \ \ \Rightarrow \ \ \ \ \ \ \ \ \ const=\frac{1}{f(0)}
\end{equation}
As a result we obtain
\[
M(x)=\frac{3m_Hm_P}{3m_H+m_P}\left[\left(\arctan \sqrt b +\frac{\ln \left(1+b\right)}{2\sqrt b}\right)\right]\times
\]
\begin{equation}
\label{eq31} \left[\left(\arctan \sqrt b +\frac{\ln \left(1+b\right)}{2\sqrt b}\right)-\left(x \arctan \sqrt b x+\frac{\ln \left(1+b x^2\right)}{2\sqrt b}\right)\right]^{-1}
\end{equation}
Our function $M(x)$ is compared with that from \cite{Aqu98} in Fig.2. It has noticeably prolonged region of constant behavior in the middle range of the interval $-1 \leq x \leq 1$.
\begin{figure}
\begin{center}
\includegraphics* [width=\textwidth] {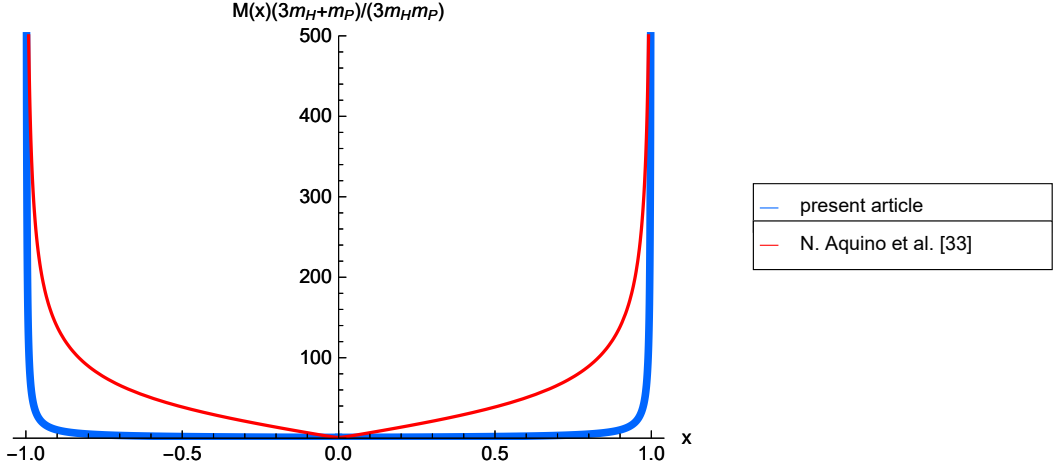}
\end{center}
\caption{The phenomenological position dependent mass (\ref{eq31}) for the inversion mode $\nu_2$ of phosphine molecule (thick line). The parameters $a = 4500$ and $b = 150$ are chosen to describe the energy levels and the splitting of the doublets for this mode in accordance with the available literature data (presented in Fig.1). The result is compared to that of \cite{Aqu98} for the position dependent mass (\ref{eq29}) (thin line).} \label{Fig.2}
\end{figure}

The available experimental values for the inversion of ${\rm{PH_3}}$ are the frequencies of $n\nu_2$ transitions $\nu_2=E_2-E_0=992.1\ {\rm cm^{-1}}$, $2\nu_2=E_4-E_0=1972.5\ {\rm cm^{-1}}$, $3\nu_2=E_6-E_0=2940.8\ {\rm cm^{-1}}$ and $4\nu_2=E_8-E_0=3895.9\ {\rm cm^{-1}}$ \cite{Spi78}, \cite{Spi86}. The reported values for the inversion barrier height are in a very wide range with $E_b=12300 \ {\rm cm^{-1}}$ \cite{Sch92} being the most plausible and often referenced value \cite{Sou15}, \cite{Sou16}. Also there are various hypothetical values for the splitting of the doublets $n\nu_2$ \cite{Sou15}, \cite{Sou16}. According to the authors of \cite{Sou15}, \cite{Sou16} the most promising region for possible detection of phosphine inversion spectrum is that of the symmetric bending band $7\nu_2$ ($\approx 6900 \ {\rm cm^{-1}}$). They predict the energy splitting for 4-th ($3\nu_2$), 8-th ($7\nu_2$) and 11-th ($10\nu_2$) doublets
\[
E_{7}-E_{6}\approx 1\cdot 10^{-5}\ {\rm cm^{-1}};E_{15}-E_{14}\approx 0.0165\ {\rm cm^{-1}};E_{21}-E_{20}\approx 7.2\ {\rm cm^{-1}}
\]
etc. The prediction of \cite{Sou16} for $3\nu_2$ region $E_{7}-E_{6}\approx 1\cdot 10^{-5}\ {\rm cm^{-1}}$ was experimentally verified in \cite{Oku18} and was not confirmed by the experiment. In the present article we consider as an example the estimate for the energy splitting of the 11-th doublet for the $10\nu_2$ band $s_{10}=E_{21}-E_{20}\approx 7.2\ {\rm cm^{-1}}$ \cite{Sou15} and derive the definite predictions for the splitting of all other doublets under the assumption that the chosen value for $E_{21}-E_{20}$ is a valid one.

Unfortunately SE with constant mass and TDWP (\ref{eq12}) fails to yield the required combination of three test values  $\{\nu_2; E_b;s_{10}\}$. Only the pair $\{\nu_2; s_{10}\}$ can be realized at $m = 1500$ and $p = 1635.8222$. For $\{\nu_2; s_{10}\}$ the achieved value of $E_b$ is $9908.5\ {\rm cm^{-1}}$. The result is depicted in Fig.1 as the dashed line. Thus our purpose is to increase $E_b$ up to the required value $E_b=12300 \ {\rm cm^{-1}}$. This is achieved by choosing the parameters $a=4500$ and $b=150$ in the position dependent mass function (\ref{eq25}). This choice adds $4452.5$ to the barrier height in the dimensionless values without touching upon the energy levels structure and considerable increase of the wings of the DWP.

We obtain the required ratios for our dimensionless energy levels $\epsilon_q$ in comparison with the literature dimensional ones $E_q$
\begin{equation}
\label{eq32} \frac{E_4-E_0}{E_2-E_0}=\frac{1972.5\ {\rm cm^{-1}}}{992.1\ {\rm cm^{-1}}}=1.98\ \ \ \ \ \ \ \ \ \ \ \ \ \frac{\epsilon_{4}-\epsilon_{0}}{\epsilon_{2}-\epsilon_{0}}=1.98
\end{equation}
\begin{equation}
\label{eq33} \frac{E_2-E_0}{E_{21}-E_{20}}=\frac{992.1\ {\rm cm^{-1}}}{7.2\ {\rm cm^{-1}}}=137.8\ \ \ \ \ \ \ \ \ \ \ \ \ \frac{\epsilon_{2}-\epsilon_{0}}{\epsilon_{21}-\epsilon_{20}}=137.8
\end{equation}
\begin{equation}
\label{eq34} \frac{E_b}{E_2-E_0}=\frac{12300\ {\rm cm^{-1}}}{992.1\ {\rm cm^{-1}}}=12.4\ \ \ \ \ \ \ \ \ \ \frac{B(m,p)+4452.5}{\epsilon_{2}-\epsilon_{0}}=12.4
\end{equation}
and satisfactorily correct ratios
\begin{equation}
\label{eq35} \frac{E_6-E_0}{E_2-E_0}=\frac{2940.8\ {\rm cm^{-1}}}{992.1\ {\rm cm^{-1}}}=2.96\ \ \ \ \ \ \ \ \ \ \ \ \ \frac{\epsilon_{6}-\epsilon_{0}}{\epsilon_{2}-\epsilon_{0}}=2.95
\end{equation}
\begin{equation}
\label{eq36} \frac{E_8-E_0}{E_2-E_0}=\frac{3895.9\ {\rm cm^{-1}}}{992.1\ {\rm cm^{-1}}}=3.92\ \ \ \ \ \ \ \ \ \ \ \ \ \frac{\epsilon_{8}-\epsilon_{0}}{\epsilon_{2}-\epsilon_{0}}=3.89
\end{equation}
The energy splitting for the 4-th doublet of the $3\nu_2$ band is $s_3=E_{7}-E_{6}=6.21\cdot 10^{-12}\ {\rm cm^{-1}}$, that for the 8-th doublet of the ($7\nu_2$) band is $s_7=E_{15}-E_{14}=0.000274\ {\rm cm^{-1}}$ and the ground ground state splitting $E_{1}-E_{0}=1.83\cdot 10^{-19}\ {\rm cm^{-1}}$. There are 15 doublets below the barrier top. At the fixed doublet $E_{21}-E_{20}\approx 7.2\ {\rm cm^{-1}}$ we obtain the strikingly smaller splitting for the low lying doublets than previously predicted in \cite{Sou16}.

We conclude that the suggested theoretical approach based on the trigonometric double-well potential is a useful tool for practitioners in IR spectroscopy. It enables one quick and convenient verifying various suppositions on the phosphine inversion spectrum. The output formulas are presented via special functions implemented in {\sl {Mathematica}}. This circumstance provides extremely easy and efficient calculation of the inversion energy levels structure for ${\rm{PH_3}}$. The Schr\"odinger equation with the position dependent mass is used for the analysis of experimental data available in literature. Our particular aim is to consider (as an example) the prediction $s_{10}=E_{21}-E_{20}\approx 7.2\ {\rm cm^{-1}}$ of Yurchenko and coauthors \cite{Sou15}, \cite{Sou16}. Fixing this doublet enables us to derive the definite prediction for the experimentally tested in \cite{Oku18} value of the energy levels splitting $E_{7}-E_{6}$ in this case. The values $\nu_2=E_2-E_0=992.1\ {\rm cm^{-1}}$, $E_b=12300 \ {\rm cm^{-1}}$ and $s_{10}=E_{21}-E_{20}\approx 7.2\ {\rm cm^{-1}}$ can be reconciled within the unified combination $\{\nu_2;E_b;s_{10}\}$ with the help of the phenomenological position dependent mass function. If the estimate from \cite{Sou15}, \cite{Sou16} $s_{10}=E_{21}-E_{20}\approx 7.2\ {\rm cm^{-1}}$ is valid then our results definitely predict for the test 4-th doublet of the $3\nu_2$ band used in the experiment of \cite{Oku18} $s_3=E_{7}-E_{6}=6.21\cdot 10^{-12}\ {\rm cm^{-1}}$ and for the ground ground state splitting $E_{1}-E_{0}=1.83\cdot 10^{-19}\ {\rm cm^{-1}}$.

\section{Appendix}
In \cite{Aqu98} the authors use the formalism of the classical mechanics to derive the position dependent mass for pyramidal molecules on the example of ${\rm{NH_3}}$. Although the classical mechanics is certainly inadequate for the description of quantum tunneling process the suggestive arguments provided by such an approach may be helpful. Here we briefly outline how our phenomenological position dependent mass function $f(x)$ given by (\ref{eq25}) can be justified within the framework of the analysis of \cite{Aqu98}. The authors of this article consider the rectangular triangle where $X$ is the dimensional distance from $N$ atom to the plane of $H$ atoms, $\rho$ is the projection of the $N-H$ distance on this plane, $\beta$ is the angle between $N-H$ and the plane of the hydrogens ($X=\rho \tan \beta$) and $r$ is $N-H$ distance ($\rho=r \cos \beta$). They derive for the position dependent mass $\mu(X)$ the expression
\begin{equation}
\label{eq37} \mu(X)=\mu_0\left[1+\frac{3m_H}{\mu_0}\left( \frac{d \rho}{dX}\right)^2\right];\ \ \ \ \ \ \ \ \ \  \ \ \ \ \ \ \ \mu_0=\frac{3m_Hm_P}{3m_H+m_P}
\end{equation}
Then they make an additional assumption that the distance $r$ is a constant $r=r_0$ during the inversion process. We release this restriction and allow an arbitrary dependence $r(X)$ or equivalently $\beta (X)$. Then
\begin{equation}
\label{eq38} \frac{d \rho}{dX}=\frac{\rho}{X}-\frac{\rho^2+X^2}{X}\frac{d\beta (X)}{dX}
\end{equation}
We want our phenomenological function $f(x)$ given by (\ref{eq25}) where $x=X/r_0$ to satisfy  $\mu(X)=M_0f(X)$, i.e.,
\begin{equation}
\label{eq39} 1+\frac{3m_H}{\mu_0}\left( \frac{d \rho}{dX}\right)^2=\frac{M_0}{\mu_0}f(X);\ \Rightarrow \ \rho (X)=\int^{X}dZ\ \sqrt{\frac{\mu_0}{3m_H}\left[\frac{M_0}{\mu_0}f(Z)-1\right]}
\end{equation}
In this particular case we obtain that $\beta (X)$ must satisfy the specific relationship ensuing from (\ref{eq38}) with $\rho$ and $d \rho/dX$ defined by (\ref{eq39}).

The pyramidal structure of ${\rm{PH_3}}$ has the bond length $\rm{PH}$ further denoted as $r_0=1.42\ \AA$ and the angle $HPH=93.5^{\circ}$ \cite{Spi78}, \cite{Spi86}.  We have the relationships of the dimensionless values with the dimensional ones
\begin{equation}
\label{eq40} L=r_0\ \ \ \ \ \ \ \ \ \ \ \ \ \ \ \ \ \ \ \ \ \ \ \ \ \ \ \ \ \ \ \ \ \ \ B(m,p)+4452.5=\frac{8M_0L^2E_b}{\hbar^2 \pi^2}
\end{equation}
As a result we have the dimensional DWP
\begin{equation}
\label{eq41} X=xL;\ \ \ \ \ \ \ \ \ \ \Omega\left(X\right)=W(X)\frac{\hbar^2}{2M_0L^2};\ \ \ \ \ \ \ \ \ \ E_q=\epsilon_q\frac{\hbar^2}{2M_0L^2}
\end{equation}
where $W(x)$ is given by (\ref{eq28}) and $\epsilon_q$ is given by (\ref{eq18}).

Finally the following methodical trick should be mentioned. One should add additional zeros in the decimal signs of the parameter $p$ at the runs of {\sl {Mathematica}} to provide the calculations of the low lying levels with the help of SF implemented in this software package.\\

Acknowledgements. The author is grateful to Prof. Yu.F. Zuev for helpful discussions. The work was supported from the government assignment for FRC Kazan Scientific Center of RAS.

\end{document}